\documentclass[iop, apj, revtex4]{emulateapj}
\usepackage{natbib}
\usepackage{url}
\usepackage{amsmath}
\usepackage{appendix}
\usepackage{hyperref}
\begin{document}

\title{Little Blue Dots in the Hubble Space Telescope Frontier Fields: Precursors to Globular Clusters?}

\author{Debra Meloy Elmegreen\altaffilmark{1} and Bruce G. Elmegreen\altaffilmark{2}
}

\altaffiltext{1}{Department of Physics \& Astronomy, Vassar College, Poughkeepsie, NY
12604; elmegreen@vassar.edu}
\altaffiltext{2}{IBM Research Division, T.J. Watson
Research Center, 1101 Kitchawan Road, Yorktown Heights, NY 10598; bge@us.ibm.com}

\begin{abstract}
Galaxies with stellar masses $<10^7\;M_\odot$ and specific star formation rates ${\rm
sSFR}>10^{-7}$ yr$^{-1}$ were examined on images of the Hubble Space Telescope Frontier
Field Parallels for Abell 2744 and MACS J0416.1-02403. They appear as unresolved
``Little Blue Dots'' (LBDs). They are less massive and have higher sSFR than
``blueberries'' studied by \cite{yang17} and higher sSFR than ``Blue Nuggets'' studied
by \cite{tacchella}. We divided the LBDs into 3 redshift bins and, for each, stacked
the B435, V606, and I814 images convolved to the same stellar point spread function
(PSF). Their radii were determined from PSF deconvolution to be $\sim80$ to $\sim180$
pc. The high sSFR suggest that their entire stellar mass has formed in only 1\% of the
local age of the universe. The sSFRs at similar epochs in local dwarf galaxies are
lower by a factor of $\sim100$. Assuming that the star formation rate is $\epsilon_{\rm
ff}M_{\rm gas}/t_{\rm ff}$ for efficiency $\epsilon_{\rm ff}$, gas mass $M_{\rm gas}$,
and free fall time, $t_{\rm ff}$, the gas mass and gas-to-star mass ratio are
determined. This ratio exceeds 1 for reasonable efficiencies, and is likely to be
$\sim5$ even with a high $\epsilon_{\rm ff}$ of 0.1. We consider whether these regions
are forming today's globular clusters. With their observed stellar masses, the maximum
likely cluster mass is $\sim10^5\;M_\odot$, but if star formation continues at the
current rate for $\sim10t_{\rm ff}\sim50$ Myr before feedback and gas exhaustion stop
it, then the maximum cluster mass could become $\sim10^6\;M_\odot$.
\end{abstract}


\section{Introduction}
\label{intro}

Large-scale deep surveys have recently enabled studies of smaller galaxies at higher
redshifts. ``Green peas" discovered from citizen science examinations \citep{card} of
the Sloan Digital Sky Survey (SDSS) are luminous compact low mass ($10^8 - 10^{10}
M_\odot$) galaxies with high specific star formation rates (sSFR), $\sim 10^{-8}$
yr$^{-1}$. They are thought to be local analogs of Lyman $\alpha$ emitters (LAEs),
which are low-mass high star formation galaxies that are increasingly common at $z >
2$. Recently, a search in SDSS for even lower mass local counterparts revealed
``blueberries,'' which are small starburst galaxies less than 1 kpc in diameter with
$\log({\rm Mass}) = 6.5$ to 7.5; they are a faint extension of the green peas
\citep{yang17}.

Here we report the discovery of even lower mass galaxies in the Hubble Space Telescope
(HST) Frontier Fields Parallels for Abell 2744 and MACS J0416.1-2403. Their properties
suggest they could be dwarf galaxies like those proposed to be the formation sites of
today's low-metallicity globular clusters \citep{elmegreen12,leaman13}. Their selection
and properties are discussed in Section \ref{dataresults} and their implications for
galaxy formation and globular clusters are in Section \ref{discussion}. A conclusion is
in Section \ref{conclusion}.

\section{Data and Results}
\label{dataresults}


The Frontier Fields comprise six galaxy clusters and six corresponding parallel fields
with deep HST imaging at optical (ACS camera) through near-infrared (WFC3 camera)
wavelengths. Archival data of two Frontier Field Parallels, Abell 2744 and
MACSJ0416.1-2403, were used for this study. Abell 2744 Parallel (heareafter A2744) and
MACSJ0416.1-2401 Parallel (heareafter M0416) have publicly available Frontier Fields
Catalogues with tabulated photometry \citep{merlin16} and photometric redshifts,
masses, and star formation rates (SFR) \citep{castellano16}. They are available through
the ASTRODEEP site\footnote{www.astrodeep.edu}. There are 3411 cataloged galaxies in
A2744 and 3732 in M0416. For our analysis we used images in the F435W, F606W, F814W and
F160W passbands (hereafter, $B, V, I, H$). The $BVI$ images reach AB mag $\sim29$ in
these two fields. They span $10800 \times 10800$ pixels; we used the mosaics with a
scale of 0.03'' pixel$^{-1}$.

\begin{figure*}
\epsscale{0.7}
\plotone{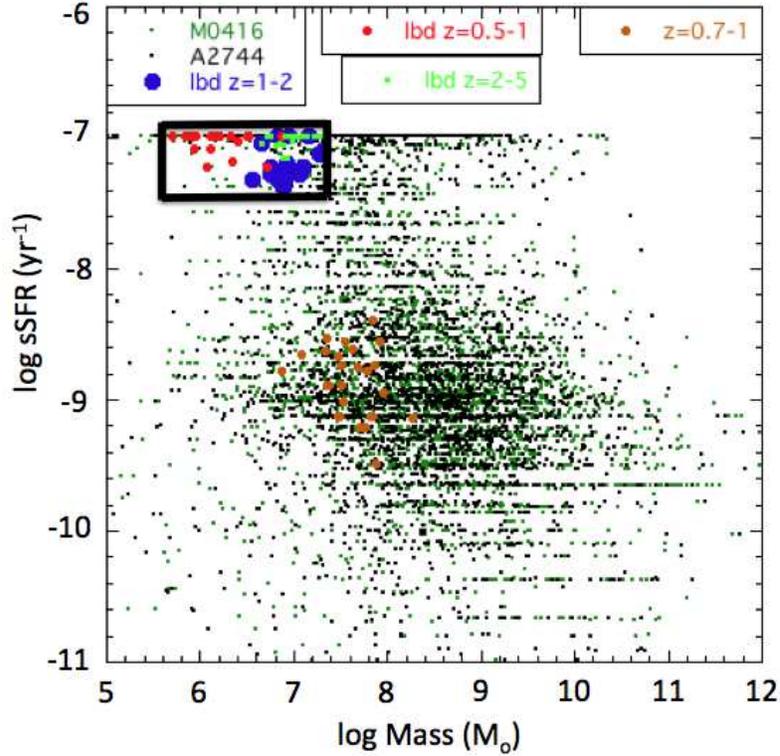}
\caption{log(sSFR /yr$^{-1}$) from catalogued SFR divided by galaxy mass is plotted versus
log(Mass) for the A2744 (green dots) and M0416 (black dots) catalogued galaxies and the
little blue dot galaxies identified in this paper (group 1 in red, group 2 in blue,
group 3 in green and group 4 in brown). The black box outlines the area completely
examined by visual inspection of the $BIH$ ASTRODEEP images. The log(sSFR) is truncated
at  -11 and log(Mass) at 12 for clarity.} \label{sSFR}
\end{figure*}

In the ASTRODEEP composite color $BIH$ images made from $B, I,$ and $H$ passbands,
several galaxies stand out as nearly point sources that are bright blue. They appear as
``Little Blue Dots'' (LBDs). We sought to examine these as examples of the low-mass,
high star formation end of the distribution of galaxies. From the Frontier Fields
catalogues of SFR and masses, we plotted specific sSFR versus mass, shown in Figure
\ref{sSFR}. There are 546 galaxies in A2744 and 547 in M0416 that have sSFR greater
than $10^{-7.4}$; of these, 173 in A2744 and 226 in M0416 have log(Mass) between 5.8
and 7.4.  We further restricted the redshift range to $0.5< z < 5.4$, resulting in 198
galaxies in A2744 and 250 in M0416. We examined all of these galaxies in the $BIH$
color images to search for LBDs. Many were eliminated either because they were too
faint to see, or were  extended galaxies rather than point sources. Many others were
eliminated because they appeared to be spurious sources. We found 55 LBDs, or about
12.2\% of this mass, redshift, and sSFR range. Considering just the mass range, the
LBDs account for 5.1\% of the tabulated galaxies.

The LBD galaxies were divided into three groups according to redshift. All have
log(sSFR) $>-7.4$. Group 1 has 19 galaxies with average $z=0.7$ and average $\log({\rm
Mass})=6.2$, group 2 has 16 galaxies with average $z=1.5$ and $\log({\rm Mass})=6.9$,
and group 3 has 20 galaxies with $z = 4$ and $\log({\rm Mass})=7.0$.  In addition, we
scanned the A2744 and M0416 images by eye to search for blue dots regardless of their
mass and sSFR. We identified 26 more small blue galaxies that were higher mass and
lower sSFR: they had average $z = 0.83$, $\log({\rm Mass}) = 7.6$, and average
$\log({\rm sSFR}) = -8.87$. These parameters place them in the normal range of sSFR
although they have a blue dot appearance; we designate them group 4. The properties of
the 4 groups are tabulated in Table \ref{tab}, to be described further below.

Figure \ref{sSFR} shows log(sSFR) as a function of log(Mass) for all of the galaxies in
the two parallel fields, A2744 with green points and M0416 with black points.  The LBD galaxies
are indicated by red, bright green, and blue points for groups 1, 2, and 3,
respectively, and brown points for group 4 galaxies that have lower sSFRs. The black
box outlines the region of completeness for identifying LBDs from a visual inspection
of low mass, high sSFR galaxies.

\begin{figure*}
\epsscale{1.0}
\plotone{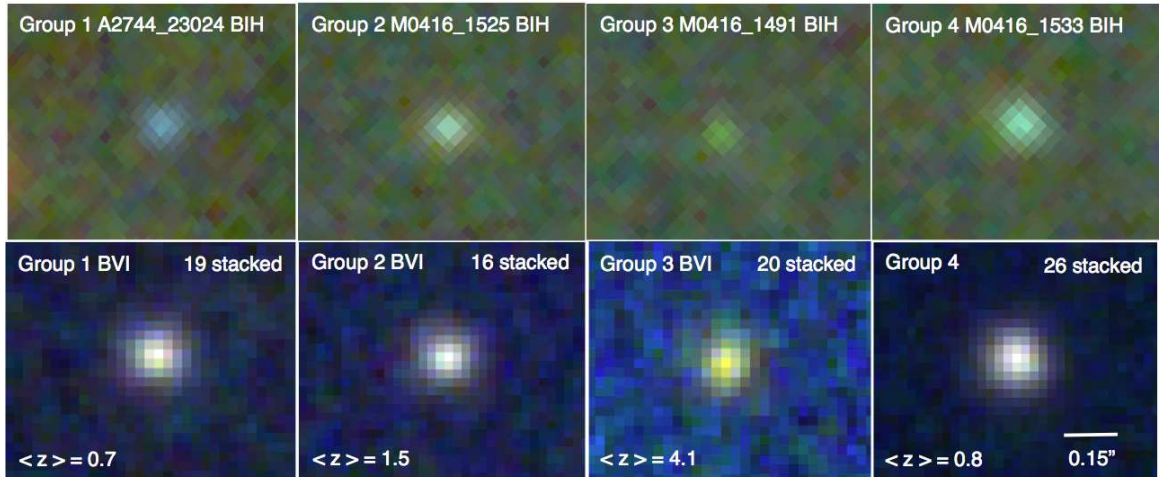}
\caption{Color images of sample galaxies and model fits. The upper panels show composite color
images from B, I, and H passbands from the ASTRODEEP site for representative galaxies
from each of the four groups listed in Table \ref{tab}. The galaxy names and fields are
listed. The lower panels are stacked images of each of the four groups, with the number
of galaxies in each stack listed along with the average redshift for each group. The
images were made by Gaussian blurring the V and I images to the B images based on the
stellar profiles. The line indicates a size of 0.15''; the scale is the same in each
image.} \label{color}
\end{figure*}

With IRAF (Image Reduction and Analysis Facility) and DS9, we made $400\times400$ pixel
cut-outs of each of the LBD galaxies for further analysis. The cut-outs were median
stacked in each filter for each group using the IRAF task {\it imcombine}, in order to
see whether they have extended faint outer disks that were not apparent in individual
galaxy images. Color $BVI$ images were made from the stacked images in each group by
first Gaussian-blurring the $V$ and $I$ images to match the $B$ images based on deconvolution with
stellar point spread function in each band. The stellar profiles were based on stacking
approximately 20 stars in each field, from \cite{thick}. The top panels of Figure
\ref{color} show composite color $BIH$ images of a sample galaxy from each of the four
groups, with the names and fields indicated. The lower panel shows composite $BVI$
stacked images for each group. The stacked images appear as nearly point sources, just
like the individual images.

\begin{figure*}
\epsscale{0.7}
\plotone{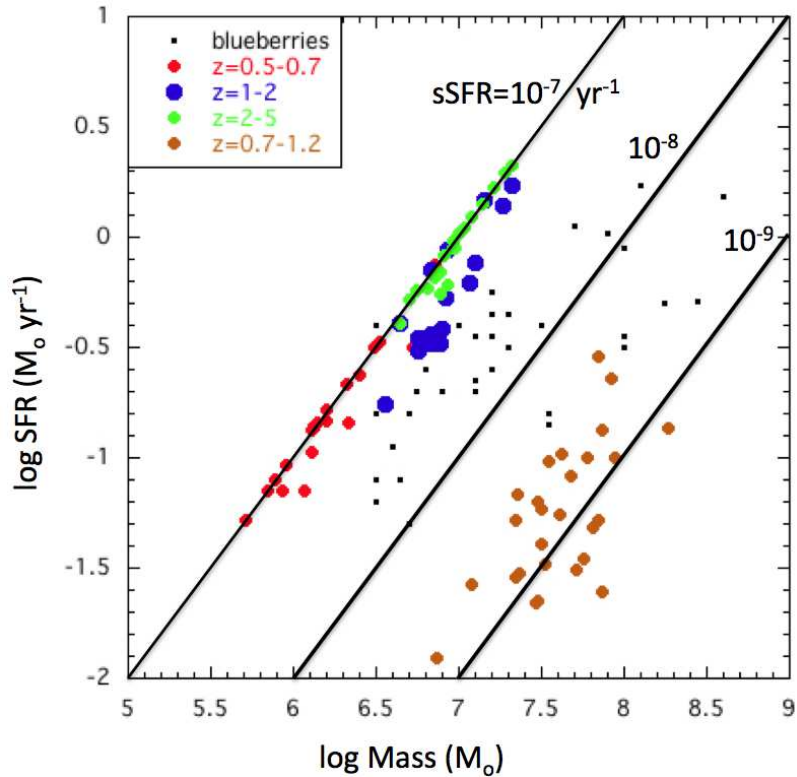}
\caption{Star formation rate in $M_\odot {\rm yr}^{-1}$ is shown as a function of mass for the four
groups of little blue dots (group 1 = red, group 2 = blue, group 3 = bright green,
group 4 = brown). The local blueberry galaxies from \cite{yang17} are shown as black dots
for comparison; they lie between the little blue dot groups. The lines indicate sSFR of
$10^{-7}$ to $10^{-9}$ yr$^{-1}$. Groups 1, 2, and 3 appear to be low mass analogs of local
blueberry galaxies. The group 4 galaxies are higher mass and have sSFR in the high end
of the normal range. Average values for each group are listed in Table \ref{tab}.}
\label{blueber}
\end{figure*}

Figure \ref{blueber} shows a $\log-\log$ plot of the SFR as a function of mass for all
four groups of LBDs as well as for the local blueberries \citep[from Figure 5
of][]{yang17}. Lines indicate constant sSFR. The group 4 galaxies (brown dots) are part
of the normal sample of galaxies in terms of their sSFR; they are at the lower-mass,
higher-sSFR limit of the SDSS galaxies in Figure 5 of \cite{yang17}. Most SDSS galaxies
have $\log({\rm sSFR}) = -10$, whereas group 4 is at the high end with $\log({\rm
sSFR}) = -9$. In contrast, the LBDs in groups 1-3 are two orders of magnitude higher in
sSFR than these, and several times higher in sSFR than the blueberries.

Group 4 galaxies also appear as LBDs on the $BIH$ images, but their sSFR are almost 2 dex
lower than the group 1-3 galaxies. Group 4 galaxies have high SFRs, making them blue,
but much more stellar mass than groups 1-3. This larger mass is indicated by their
$(B-H)$ color index, which is several tenths of a magnitude redder than that of groups
1-3 galaxies. Average masses and $(B-H)$ indices are listed in Table \ref{tab}.

Gaussian fits were done for each stacked image in each filter for the four groups. The
resulting dispersions, $\sigma$, were deconvolved from the stellar dispersions in order
to get the average galaxy radii (assumed to be the half widths at half maxima,
$2.35\sigma$) in each passband. Fractional uncertainties in the deconvolved radii were
determined from the quadratic sums of the fractional uncertainties in the galaxy and
stellar sizes. For groups 1, 2, and 4, we used $\sigma$ from the I-band images because
they were the highest quality. For group 3, which has the highest redshift, we used the
V image. The angular size was converted to linear size for each redshift using a
$\Lambda$CDM model \citep[][$\Omega_{\rm m}=0.315$, $H= 67.3$ km s$^{-1}$
Mpc$^{-1}$]{ade14}. The results are in Table \ref{tab}.

\section{Discussion}
\label{discussion}

\subsection{A Recent Burst}

The LBD galaxies have a high sSFR, on the order of $10^{-7}$ yr$^{-1}$. From the
$\Lambda$CDM model, we determined the age of the Universe corresponding to the average
redshift of each LBD group, and multiplied that by the average sSFR for that group. The
values in Table 2 range from $\sim147$ to $\sim513$ for groups 1-3, but it is only
$\sim9$ for group 4. The inverse of these numbers is the fraction of the age of the
universe during which the observed stellar mass has been formed at the observed SFR.
The very high values suggest that star formation in group 1-3 LBDs began in the last
one percent or less of the local age of the Universe.

\subsection{Gas Mass Estimates}
\label{gasmass}

We estimate the gas mass in the LBD galaxies using the usual relation for the SFR, $S$:
\begin{equation}
S=\epsilon_{\rm ff}M_{\rm gas}/t_{\rm ff}\label{eq:S}
\end{equation}
for efficiency per free-fall time $\epsilon_{\rm ff}$ \citep[e.g.][]{krumholz05}, gas
mass $M_{\rm gas}$ and free fall time $t_{\rm ff}=(32G\rho/[3\pi])^{-1/2}$ for gas
density $\rho=M_{\rm gas}/(4\pi R^3/3)$; $R$ is the deconvolved radius from the stacked
images. After re-arranging terms,
\begin{equation}
M_{\rm gas}=\left(\frac{\pi^2}{8G}\right)^{1/3}R\left(S/\epsilon_{\rm ff}\right)^{2/3}.
\end{equation}
Because $\epsilon_{\rm ff}$ is not measured, we evaluate $M_{\rm gas}\epsilon^{2/3}$
from the observed quantities,
\begin{equation}
M_{\rm gas}\epsilon^{2/3}=6.5\times10^4 R\times S^{2/3} \;M_\odot
\end{equation}
where now $R$ is in parsecs and $S$ is in $M_\odot$ yr$^{-1}$.

The results are in Table 2: $\log(M_{\rm gas}\epsilon_{\rm ff}^{2/3})\sim6.7$ for
groups 1, 2, and 3, and $\sim6.5$ for group 4, which are more like normal galaxies. The
ratio of this efficiency-normalized gas mass to the observed stellar mass is also
given, and is of order unity for groups 1-3.

These gas-to-star ratios are unusually high for galaxies in groups 1-3, considering
that the usual Kennicutt-Schmidt relation has $\epsilon_{\rm ff}\sim0.01$, which would
make the gas-to-star ratios 45, 16, and 12, respectively. Because an efficiency of
$\epsilon_{\rm ff}\sim1$ is difficult to justify (i.e., it would imply that all of the
gas converts to stars in one free-fall time, without feedback, low-density gas, etc.),
the gas-to-star ratio is more likely $\sim5$, which is what $\epsilon_{\rm ff}\sim0.1$
would give.

The sSFR in LBDs is also extreme in another sense. The inverse, $\sim10^7$ years,
implies that nearly all of the stars have formed within the last 10 Myr, which is too
short a time for most of the massive stars to have supernovaed. This suggests a high
star formation efficiency and the formation of bound clusters \citep{parmentier12}.

\subsection{Star Formation Histories of Local dIrrs}

Star formation histories of local dwarf galaxies do not show an early phase with a SFR
as high as in the LBDs, which is $\sim0.4\;M_\odot$ yr$^{-1}$ from the average of the
$\log({\rm SFR})$ for groups 1-3 in Table 1.

In the local dIrr WLM, the peak of the SFR $9-12$ Gyr ago was $\sim
6.7\pm1.5\times10^{-4}\;M_\odot$ yr$^{-1}$ \citep{dolphin00a} when the stellar mass was
slightly less than in the LBDs, $\sim 8 \times10^5\;M_\odot$ \citep{leaman12a}. This
makes $\log({\rm sSFR})\sim-9.1$ for sSFR in yr$^{-1}$, two orders of magnitude lower
than for LBDs. At the average redshift of $z=0.73$ for group 1, the age of the forming
stars today would be 6.6 Gyr. In WLM, the SFR at around 6.6 Gyr ago was $\sim
1.3\pm1.5\times10^{-4}\;M_\odot$ yr$^{-1}$ \citep{dolphin00a}, and the stellar mass was
$\sim4\times10^6\;M_\odot$ (from an integral over the SFR), which is a similar $M_{\rm
star}$ to that in the LBDs at that age (Table 1). However, the $\log({\rm sSFR})$ in
WLM was lower by more than three orders of magnitude, $\sim-10.5$ for the same units.

For the local dSph galaxy Fornax, \cite{deboer12a} determined a SFR that averages
$\sim1.6\pm0.5\times10^{-3}$ between 10 and 14 Gyr ago, and builds up a stellar mass of
$6.5\times10^6\;M_\odot$ by the end of that time. Thus $\log({\rm sSFR})\sim-9.6$. At
$5-10$ Gyr ago, the SFR was $3.3\pm0.3\times10^{-3}\;M_\odot$ yr$^{-1}$, and $M_{\rm
star}\sim1\times10^7\;M_\odot$, so $\log({\rm sSFR})\sim-9.5$.

Another local dSph galaxy, Sculptor, has more of a starburst than Fornax at early times
\citep{deboer12b}. The last significant event was $8-10$ Gyr ago inside $\sim270$ pc.
In the period from $12-14$ Gyr ago, the SFR was $\sim2.4\times10^{-3}\;M_\odot$
yr$^{-1}$ and $M_{\rm star}\sim4.8\times10^6\;M_\odot$, making $\log(sSFR)=-9.3$ over a
radius of 1.5 kpc. At $8-10$ Gyr ago, the SFR was $\sim0.45\times10^{-3}\;M_\odot$
yr$^{-1}$ and $M_{\rm star}\sim7.6\times10^6\;M_\odot$, so $\log(sSFR)=-10.2$ mostly
inside $\sim400$ pc. This is about the radius of the LBDs which have $\sim10^3\times$
higher sSFR.

Carina is a third local dSph studied by the same group \citep{deboer14}.  Between
$10-14$ Gyr, the SFR was $\sim5\times10^{-5}\;M_\odot$ yr$^{-1}$ and the accumulated
$M_{\rm star}\sim2\times10^5\;M_\odot$, making $\log(sSFR)=-9.6$ in yr$^{-1}$. Near the
SFR peak at 5 Gyr ago, the SFR was $\sim1.3\times10^{-4}\;M_\odot$ yr$^{-1}$ and the
accumulated $M_{\rm star}\sim7\times10^5\;M_\odot$, so $\log(sSFR)=-9.7$ in yr$^{-1}$.

In these examples the whole-galaxy sSFR is 2 to 3 orders of magnitude lower than what
is observed in the LBDs for the same epoch in the Universe. Perhaps the age resolution
in local galaxies does not show a short-lived SFR burst that is as high as that in the
LBDs.

A further comparison is with the early SFRs for the globular clusters in the Fornax
dSph galaxy, as determined by \cite{deboer16}. From their figure 8, the peak rates for
Fornax globular clusters 1-5 are, in units of $10^{-5}\;M_\odot$ yr$^{-1}$,
approximately: $2$, $6$, $16$, $2$, and $4.5$, with an rms of about 1 to 2 in these
units. The cluster masses from their table 3 are, in units of $10^5\;M_\odot$,
$0.42\pm0.10$, $1.54\pm0.28$, $4.98\pm0.84$, $0.76\pm0.15$, and $1.86\pm0.24$,
respectively. The average of the logs of the ratios of the SFRs to the masses of the
clusters is $\log{\rm sSFR}=-9.5$, which is still much smaller than the sSFR we observe
here. This star formation is spread out over $\sim2$ Gyr in their figure 8, however,
and perhaps much of this spread is from age uncertainties. If the SFR were 100 times
more concentrated in time, taking place in only 20 Myr for each globular cluster, then
the sSFRs would be comparable. 20 Myr is not an unreasonable timescale for cluster
formation.

\subsection{Globular Cluster Formation}

Star-forming regions produce a fraction, $\Gamma$, of their total stellar mass in the
form of bound clusters, and these clusters form with a mass distribution function that
spreads out the total into many individual clusters, most of which will not survive for
a Hubble time. The mass distribution is often expressed as a Schechter function
\citep{adamo15} with a power law slope of $-2$ (for linear intervals of mass) and an
exponential drop at some high cluster mass, $M_{\rm c}$. For high SFR, $M_{\rm c}$ can
be larger than $10^6\;M_\odot$ \citep[e.g.,][]{zhang99} and the distribution function
is essentially a power law up to a globular cluster mass. Considering this limit, or
the case of a pure power law \citep{whitmore14}, integrals over the distribution
function are analytic, and the total mass of stars that accompanies a single massive
cluster with $M_{\rm GC}=10^6\;M_\odot$ is:
\begin{equation}
M_{\rm star}=\Gamma^{-1}M_{\rm GC}\left(1+\ln[M_{\rm GC}/M_{\rm min}]\right)
\end{equation}
where $M_{\rm min}$ is the minimum cluster mass \citep{elmegreen12}. Setting
$\Gamma\sim0.25$ \citep{chandar17} and $M_{\rm min}\sim100\;M_\odot$, we derive $M_{\rm
star}=4\times10^7\;M_\odot$.  At the SFRs of the LBDs, which average $\sim0.4\;M_\odot$
yr$^{-1}$ as mentioned above, it will take $\sim100$ Myr to make enough stars to have a
single $10^6\;M_\odot$ cluster.  At the current average $M_{\rm star}=10^{6.7}$ from
Groups 1-3 in Table 1, the largest cluster mass is likely to be
$1.5\times10^5\;M_\odot$.

The observed starbursts could continue in these young regions for several tens of Myr.
Usually, star formation continues in a region for a few dynamical times, perhaps up to
$10t_{\rm ff}$ at the average density \citep{elmegreen00}. Feedback and gas exhaustion
seem to stop regions from continuing much longer than this, except perhaps in dispersed
cloud fragments.  Using the gas masses from Table 2 with $\epsilon_{\rm ff}\sim0.1$
discussed above, and using the average radii for the deconvolved LBDs, we derive the
average gas densities and free-fall times listed in Table 2.  The average for groups
1-3 is $t_{\rm ff}\sim5\times10^6$ yr. If star formation continues for $\sim10t_{\rm
ff}$ at the average SFR of $\sim0.4\;M_\odot$ yr$^{-1}$, then the final stellar mass in
the burst will be $M_{\rm star}\sim2\times10^7\;M_\odot$ and the maximum cluster mass
is likely to be $5.2\times10^5$. If $\Gamma=0.5$ \citep{adamo11} for the SFR surface
densities of LBDs, which average $\sim16\;M_\odot$ yr$^{-1}$ kpc$^{-2}$ for groups 1-3,
then this maximum expected mass is $10^6\;M_\odot$.

\cite{zaritsky16} examined the globular cluster number as a function of galaxy mass for
local galaxies. From their figure 4, they derived a cluster mass fraction of 0.013 for
$M_{\rm star}=10^{8.5}\;M_\odot$. These globular clusters are typically older than most
of a galaxy's stars, so they were presumably present when the galaxy had only 10\% of
its present mass. If we consider $0.1\times$ the present galaxy mass and $10\times$ the
present globular cluster mass to account for cluster mass loss according to current
models for the origin of multiple stellar populations in these clusters
\citep{dec07,dercole08,webb15}, then the mass fraction in globular clusters at that
early time was $0.013\times10/(0.013\times10+0.1)=0.57$.  Thus young versions of
galaxies like these could have been dominated in mass by one or two globular clusters
when the clusters formed. This situation is similar to what we may be seeing with the
LBDs.

\subsection{Evidence from Simulations}

Cosmological simulations indicate that star-forming galaxies go through stages of
compaction. \cite{tacchella} found that compact high SFR galaxies have high gas
fractions with short depletion timescales; they referred to these galaxies as ``blue
nuggets''. Their Figure 2 shows sSFR versus mass for their proposed evolutionary
sequence of high sSFR galaxies evolving into quenched galaxies. The lower stellar mass
limit in their simulations is $\log M_{\rm star}= 7$, which is approximately the mass
of a LBD galaxy. Their $z \sim 6$ galaxies that start out as low-mass, high-sSFR
galaxies become high-mass, low-sSFR galaxies by $z = 1$. The sSFR for the high redshift
galaxies is $\log({\rm sSFR}/{\rm yr}^{-1}) = -8.5$. These simulated galaxies are not
as extreme in sSFR as the LBDs, which at $\log M_{\rm star}=7$ have $\log({\rm
sSFR}/{\rm yr}^{-1}) = -7$.

Zoom-in simulations of galaxy formation in a cosmological context were used to study
cluster formation in \cite{pfeffer17} using a detailed sub-grid prescription. The
galaxies readily formed massive clusters because of their high ambient pressures. LBDs
are lower mass galaxies than they studied, but LBDs could still have relatively high
pressures considering the high gas densities in Table 2.   Observations of
gravitationally lensed massive star-forming regions that could contain globular
clusters progenitors were in \cite{vanzella17}.

\section{Conclusions}
\label{conclusion}

Low mass galaxies ($\log(M_{\rm star}/M_\odot)<7.4$) with high sSFR ($\log (sSFR/{\rm
yr}^{-1})>-7.4$ in two Frontier Field Parallels were examined by eye and found to have
a characteristic appearance which we have termed Little Blue Dots (LBDs). A more
complete survey of these fields uncovered more LBDs of higher mass. The low mass LBDs
have such high sSFR that they appear to have formed all of their stars in the last 1\%
of the age of the universe for them. They appear to be gas-dominated compared to stars,
perhaps by a factor of 5, and midway through the process of forming massive clusters
that will eventually be the globular clusters of today. These clusters would have
represented a high fraction of the stellar mass in these systems when they formed, and
that high fraction is consistent with the observed mass fraction in local dwarf
galaxies. We suggest that objects like this are the long-sought progenitors of
low-metallicity globular clusters, which formed in dwarf galaxies and got assimilated
into the halos of today's spirals and ellipticals.

{\it Acknowledgments} We thank Marc Rafelski for his helpful discussions about stacking
images.

\begin{deluxetable}{lccccccc}
\tabletypesize{\scriptsize}\tablecolumns{8} \tablewidth{0pt} \tablecaption{Little Blue Dot Galaxies: Stack Results}
\tablehead{\colhead{Group}&
\colhead{no.}&
\colhead{$z$}&
\colhead{log(Mass)} &
\colhead{Radius} &
\colhead{$(B-H$)}&
\colhead{log(SFR)}&
\colhead{log(sSFR)}\\
\colhead{}&
\colhead{}&
\colhead{}&
\colhead{$M_{\odot}$}&
\colhead{pc}&
\colhead{mag}&
\colhead{$M_{\odot} {\rm yr}^{-1}$}&
\colhead{yr$^{-1}$}
}
\startdata

1&19&$0.73\pm0.11$	&$6.20\pm0.29$	&	$183\pm69$	&$0.16\pm0.76$&$-0.83\pm0.29$&$-7.03\pm0.08$\\
2&16&$1.45\pm0.28$	&$6.92\pm0.21$	&	$148\pm28$	&$0.55\pm0.41$&$-0.27\pm0.28$&$-7.19\pm0.13$\\
3&20&$4.09\pm0.86$	&$6.97\pm0.19$	&	$84\pm38$	&$0.69\pm0.55$&$-0.05\pm0.21$&$-7.02\pm0.05$\\
4&26&$0.83\pm0.09$	&$7.61\pm0.30$	&	$331\pm77$	&$1.02\pm0.64$&$-1.26\pm0.33$&$-8.87\pm0.27$\\

\enddata
\label{tab}
\end{deluxetable}

\begin{deluxetable}{lccccc}
\tabletypesize{\scriptsize}\tablecolumns{8} \tablewidth{0pt} \tablecaption{Derived Quantities}
\tablehead{
\colhead{Group}&
\colhead{$sSFR\times$}&
\colhead{$\log(M_{\rm gas}\epsilon_{\rm ff}^{2/3})$}&
\colhead{$M_{\rm gas}\epsilon_{\rm ff}^{2/3}/M_{\rm star}$}&
\colhead{Avg. Density}&
\colhead{$\log(t_{\rm ff})$}\\
\colhead{}&
\colhead{age of Univ.}&
\colhead{$M_\odot$}&
\colhead{}&
\colhead{atoms cm$^{-3}$}&
\colhead{yr}
}

\startdata

1&469&6.5&2.1&19&7.0\\
2&513&6.8&0.76&68&6.7\\
3&147&6.7&0.54&290&6.4\\
4&9.0&6.5&0.076&2.9&7.4
\enddata
\end{deluxetable}

\newpage

\newpage

\newpage

\end{document}